\documentclass[PRL,showpacs,amsmath,amssymb,superscriptaddress,twocolumn]{revtex4-1}

\usepackage[T1]{fontenc}
\usepackage[latin1]{inputenc}
\usepackage{graphicx}
\usepackage{epsfig}
\usepackage{dcolumn}
\usepackage{bm}
\usepackage{bbm}
\usepackage{dsfont}
\usepackage{amssymb,amsmath}
\usepackage{amsfonts}
\usepackage{float}
\usepackage{upgreek}
\usepackage{mathrsfs}
\usepackage{color}

\newcommand{\beq}{\begin{equation}}
\newcommand{\eq}{\end{equation}}
\newcommand{\bea}{\begin{eqnarray}}
\newcommand{\ea}{\end{eqnarray}}

\newcommand{\nn}{\nonumber}

\begin{document}


\title{The interplay of phonon and exciton-mediated superconductivity in hybrid semiconductor-superconductor structures}
\author{Petros Skopelitis}
\email[Email: ]{zvap078@live.rhul.ac.uk}

\affiliation{Department of Physics, Royal Holloway, University of London, Egham, Surrey TW20 0EX, United Kingdom}
\affiliation{Department of Physics and Astronomy, University of Southampton, Highfield, Southampton SO17 1BJ, United Kingdom}

\author{Evgenia D. Cherotchenko}
\email[Email: ]{E.Cherotchenko@soton.ac.uk}
\affiliation{Department of Physics and Astronomy, University of Southampton, Highfield, Southampton SO17 1BJ, United Kingdom} 
\affiliation{ITMO University, Saint Petersburg 197101, Russia}

\author{Alexey V. Kavokin}
\email[Email: ]{A.Kavokin@soton.ac.uk}

\affiliation{Department of Physics and Astronomy, University of Southampton, Highfield, Southampton SO17 1BJ, United Kingdom} 
\affiliation{CNR-SPIN, Viale del Politecnico 1, I-00133, Rome, Italy}

\author{Anna Posazhennikova}
\email[Email: ]{anna.posazhennikova@rhul.ac.uk}

\affiliation{Department of Physics, Royal Holloway, University of London, Egham, Surrey TW20 0EX, United Kingdom}

\date{\today}

\begin{abstract}
We predict a strong enhancement of the critical temperature in a conventional Bardeen-Cooper-Schrieffer (BCS) superconductor in the presence of a bosonic condensate of exciton-polaritons. The effect depends strongly on the ratio of the cutoff frequencies for phonon and exciton-polariton mediated BCS superconductivity, respectively. We also discuss a possible design of hybrid semiconductor-superconductor structures suitable for the experimental observation of such an effect.
\end{abstract}


\maketitle



{\it Introduction}  
There  have  been  enormous  efforts  to  realise superconductivity at  higher  temperatures, especially in a form similar to BCS superconductivity \cite{BCS1957}, which involves the formation of Cooper pairs. In the search of high $T_c$ superconductivity, it is generally agreed that there are two main ways to achieve high $T_c$: (a) by discovering or creating a system where the mediators (phonons or other excitations) of Cooper pairing have high characteristic energies (higher then the typical Debye scale $\omega_D$ found in BCS metals) and (b) by increasing the coupling strength of the mediators with electrons \cite{Cotlet2016}. However, increasing the coupling strength may lead to lattice instabilities \cite{Allen1982}, and materials with higher Debye energy do not necessarily have larger coupling constant. In this context, since 1970s the special attention has been paid to the out-of-thermal equilibrium systems, where the strength of electron-electron coupling may be mediated by crystal excitations other than phonons. In particular, a lot of works were devoted to the exciton mediated superconductivity \cite{ABB1973,Ginzburg1976}. While there is no unambiguous experimental evidence for the exciton-mediated superconductivity reported till now,  recently, the similar phenomenon of light-induced superconductivity has been discovered \cite{Fausti2011,Cavalleri2016}. In these experiments, light serves for generation of crystal excitations similar to excitons that help electron-electron pairing. 

In the last decade, several theoretical proposals on the superconductivity mediated by a Bose-Einstein condensate of excitons (exciton-polaritons) have been published \cite{laussy2010exciton,Cherotchenko2016, Cotlet2016}. These proposals are based on a tremendous progress in the experimental studies of bosonic condensates of exciton polaritons at elevated temperatures \cite{Christopoulos2007}. These studies pave wave to observation of superconductivity in semiconductor structures under optical pumping. 

While the light- or exciton-mediated superconductivity is in the focus of interest now, it is yet far from being clear what kind of material system would be the most suitable for the observation of such phenomena, especially at high temperatures. In the present work we show a high potentiality of hybrid superconductor-semiconductor systems, where the interplay of a conventional phonon-mediated BCS and the superconductivity mediated by an excitonic condensate may lead to a sharp increase of $T_c$. Recent experiments \cite{Cavalleri2016} also indicate higher superconducting temperatures when superconductivity is light-induced. Our setup is, however, very different to experimental systems of \cite{Fausti2011,Cavalleri2016}, neither do we consider short-time superconductivity as in these experiments.

\begin{figure}[!bt]
\begin{center}
\includegraphics[width=0.5\textwidth]{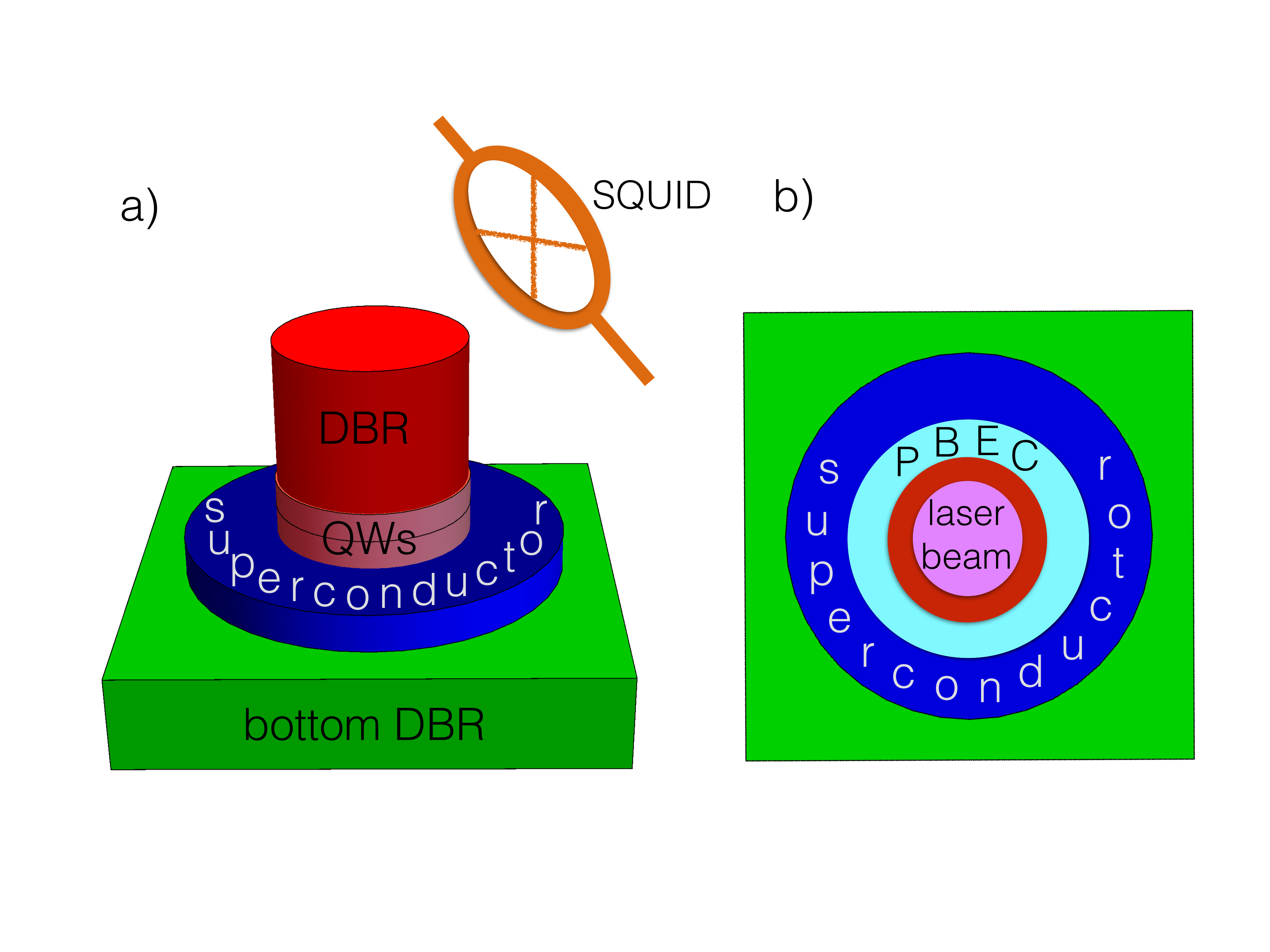}
\end{center}
\caption{The diagram of a structure suitable for observation of the interplay of phonon- and exciton (bogolon)-induced superconductivity. A superconducting ring is deposited around pillar semiconductor microcavity. DBRs denote distributed Bragg reflectors. The condensate of polaritons is excited by a laser (a). View from the top (b). }
\label{setup}
\end{figure}

We develop a simple model illustrating how our mechanism of achieving high $T_c$  would work.  We also propose a specific experimental set-up with a superconductor (SC) --- quantum wells (QWs) heterostructure embedded in a semiconductor microcavity shown in Fig. \ref{setup}.  In such a setup the combined effect of the phonon coupling in conventional BCS superconductors and the light-induced electron-electron coupling mediated by a bosonic condensate of exciton-polaritons should be realized. We revisit the Bose-Fermi system considered in \cite{laussy2010exciton, laussy2012polariton}, but take into account two types of bosonic excitations instead of one: the "fast" bogolons resulting from density fluctuations of the polaritons in the polaritonic Bose-Einstein condensate (pBEC) \cite{laussy2010exciton, laussy2012polariton}, and the "slow" acoustic phonons of the metal plate. By generalizing Gor'kov equations \cite{gor1958energy} for this case, we derive critical temperature which can be high due to interference effects of the two interactions at long distances and strongly depends on relative sizes of characteristic cut-off frequencies for phonons and bogolons (the excitations of the pBEC).

There are several significant advantages of our setup, where superconductivity is assisted by light, over typical suggestions from the past \cite{ABB1973}. The seminal work \cite{ABB1973} is based on the  original Ginsburg ideas of excitonic superconducting mechanism \cite{Ginzburg1971}.
A thin metallic layer on a semiconductor surface was suggested as a possible experimental setup for realization of excitonic mediated coupling between electrons \cite{ABB1973}. It was shown that in order to see any results in $T_c$ the excitonic coupling constant $\lambda_{ex}$ should be at least of the order of 0.2 or 0.3. These values of $\lambda_{ex}$ turned out to be very challenging from experimental point of view and have been still not achieved  (see also Ref.\cite{Rhim2016} and references therein). 

In our setup the BCS coupling constant  and therefore the "bare" critical temperature $T_{c0}$ of the SC are given (along with its characteristic Debye frequency $\omega_D$). The great advantage of the exciton-polariton induced coupling is that the coupling strength can be controlled experimentally, e.g. it was shown the coupling is proportional to polariton density \cite{laussy2010exciton, laussy2012polariton}. Moreover, the cut-off frequency of the polaritons $\omega_B$ can be also controlled and is defined by the microcavities properties. As we demonstrate below the control of the two parameters can lead to a notable increase of $T_{c}$ in comparison with $T_{c0}$. In our structure any increase of the measured $T_c$ in comparison with the reference temperature $T_{c0}$ will confirm the interplay of the two coupling mechanisms. 

{\it Model Hamiltonian} We develop a simple model for the setup in Fig. \ref{setup}. The setup comprises a semiconductor microcavity with embedded QWs and a 2D layer of a conventional SC separated from the wells by a thin barrier. The bosonic condensate of exciton-polaritons is generated by a continuos wave (CW) pulse in the quasi-stationary regime,  which is a well-established technique nowadays \cite{Kasprzak2006,Lagoudakis2008,Levrat2010}. Its density $N_0$ can be controlled by the pump intensity. The diameter of the total structure could be around $50 \mu m$ or less with the micropillar diameter of $20-30 \mu m$.

Importantly, the suggested experimental geometry and structure design allow for the strong reduction of light absorption in the superconducting ring. The light absorption usually leads to an unwanted increase of the effective temperature of the electron gas and therefore hampers the experiments. We chose the quasi-2D geometry since electron-exciton interaction that is crucial for the exciton-mediated superconductivity is then maximized. The hamiltonian reads
\begin{equation}
    H=H^0_e+H^0_{p}+H_{e-e}+H_{e-p}+H_{p-p}+H_{e-ph},
    \label{Ham}
\end{equation}
where $H_e^0, H_{p}^0$ are the electron and polariton kinetic terms
\bea
H^0_e=\int \psi^\dagger_{\alpha}({\bf x})\biggl\{-\frac{1}{2m_e}\nabla^2-\upmu_e\biggr\}\psi_{\alpha}({\bf x})d{\bf x}, \nn \\
 H^0_{p}=\int \phi^\dagger({\bf R})\biggl\{-\frac{1}{2m_p}\nabla^2-\upmu_p\biggr\}\phi({\bf R})d{\bf R}.
\label{kin}
\ea
Here the electron field operators are
\bea
\psi_\alpha({\bf x})=\frac{1}{\sqrt{A}}\sum_{{\bf k}}\Psi_{k,\alpha}({\bf x})c_k=
\frac{1}{\sqrt{A}}\sum_{{\bf k}}e^{i{\bf k}\cdot{\bf x}}\eta_a c_{{\bf k}}, \nn \\
 \psi^\dagger_\alpha({\bf x})=\frac{1}{\sqrt{A}}\sum_{{\bf k}}\Psi^*_{k,\alpha}({\bf x})c^\dagger_k=\frac{1}{\sqrt{A}}\sum_{{\bf k}}e^{-i{\bf k}\cdot{\bf x}}\eta_ac^\dagger_{{\bf k}},
\label{elec_field}
\ea
where $\eta_a$ are the two spin functions, and $c_{\bf k}, c^\dagger_{{\bf k}}$ are fermionic creation and annihilation operators, $A$ is the area of the metallic plate. Polaritons are described by the field operators 
\begin{equation}
   \begin{split}
       \phi({\bf R})&=\frac{i}{\sqrt{A}}\sum_{{\bf P}}e^{i{\bf P}\cdot{\bf R}}b_{{\bf P}},\\
       \phi^\dagger({\bf R})&=\frac{i}{\sqrt{A}}\sum_{{\bf P}}e^{-i{\bf P}\cdot{\bf R}}b^\dagger_{{\bf P}},
   \end{split} 
\end{equation}
where ${\bf P}$ and ${\bf R}$ are the polariton's centre of mass momentum and the polariton's centre of mass coordinate, and $b_{{\bf P}}$ and $b^\dagger_{{\bf P}}$ are bosonic annihilation and creation operators. 

The interaction terms in Eq. \eqref{Ham} include electron-electron interaction
\beq
H_{e-e}=\int\psi^\dagger_{\alpha}({\bf x})\psi^\dagger_{\beta}({\bf x'})V_c({\bf x}-{\bf x'})\psi_{\alpha}({\bf x})\psi_{\beta}({\bf x'})d{\bf x} d{\bf x'},
\eq
electron-polariton interaction
\beq
H_{e-p}=\int\psi^\dagger_{\alpha}({\bf x})\psi_{\alpha}({\bf x})V_{e-p}({\bf x}-{\bf R})\phi^\dagger({\bf R})\phi({\bf R})d{\bf x} d{\bf R},
\eq
interaction between polaritons
\beq
H_{p-p}=\int\phi^\dagger({\bf R})\phi({\bf R})V_{p-p}({\bf R}-{\bf R'})\phi^\dagger({\bf R'})\phi({\bf R'})d{\bf R} d{\bf R'},
\eq
and electron-phonon interaction
\beq
H_{e-ph}=-e\int\psi^\dagger_{\alpha}({\bf x})\psi_{\alpha}({\bf x})V_{e-ph}({\bf x}-{\bf x'})\rho({\bf x'})d{\bf x} d{\bf x'}.
\eq
 Here $V_c({\bf  x}-{\bf x})$ is the screened Coulomb repulsion potential,   $V_{p-p}({\bf R}-{\bf R})$ is contact interaction between the exciton- polaritons, electron-polariton $V_{e-p}({\bf x}-{\bf R})$ and electron-phonon $V_{e-ph}({\bf x}-{\bf x'})$ potentials can be also taken as contact (see Supplemental Material), $\rho({\bf x})$ in the background surface charge density of the lattice. 

After performing the standard Bogoliubov transformation on the exciton-polariton condensate we arrive to the effective hamiltonian, which takes into account the interaction of electrons with {\it bogolons} (elementary excitations of the exciton-polariton condensate)
\beq
H=H_e^0+H_{e-e}+H_{e-bog}+H_{e-ph}.
\eq
Here we have
\bea
     H_{e-bog}=\gamma_1\int\psi^\dagger_\alpha({\bf x})\psi_\alpha({\bf x})\phi_1({\bf x})d{\bf x}, \nn \\
    H_{e-ph}=\gamma_2\int\psi^\dagger_\alpha({\bf x})\psi_\alpha({\bf x})\phi_2({\bf x})d{\bf x},
    \label{bog_ph}
\ea
where $\phi_1({\bf x})$ is the bosonic field operator of bogolons, and $\phi_2({\bf x})$ is the field operator of phonons, $\gamma_{1(2)}$ is the electron-bogolon (electron-phonon) coupling constant (see Supplemental Material). 

{\it Effective attraction and gap equation} We now study the system of electrons with the effective bogolon- and phonon-mediated attractions (the Coulomb interaction is neglected for the moment). The effective hamiltonian will then have two contributions ($j=1,2$)
\beq
H_{e-e}^{eff}=-\frac{V_j}{2}\int d{\bf{x}} \psi^\dagger_{\alpha}({\bf x})\psi^\dagger_{\beta}({\bf x})\psi_{\beta}({\bf x})\psi_{\alpha}({\bf x}),
\label{ee_eff}
\eq
where $V_1=(\gamma_1^2+\gamma_2^2)$ and $V_2=\gamma_2^2$ have different ranges in momentum space: $V_1$ is constant for $|\xi'-\xi|<\omega_B$, and zero otherwise, while $V_2$ is constant for $\hbar\omega_B<|\xi'-\xi|<\hbar\omega_D$ (with $\xi, \xi'$ being energies counted from the Fermi energy as usual in the BCS theory). Here we assumed the inequality $\omega_B<\omega_D$ for the following reason:  the characteristic energy cut-off for polaritonic condensates is expected to be of the order of 100 K in high quality inorganic microcavities, determined by the Rabi splitting (which is tuneable and depends on the microcavity parameters). For a conventional weak-coupling superconductor (e.g. Al) we expect the Debye energy to be of the order of 400K, which is larger than $\hbar\omega_B$ and much larger than $k_BT_c\approx 1K$. 

In the mean-field theory the hamiltonian \eqref{ee_eff} becomes
\bea
H_{e-e}^{eff}&=&-\int d{\bf x}\,\biggl\{\Delta^*_j({\bf x})\psi_{\uparrow}({\bf x})\psi_{\downarrow}({\bf x})
 \nn \\ &+&\psi^\dagger_{\downarrow}({\bf x})\psi^\dagger_{\uparrow}({\bf x})\Delta_j({\bf x})\biggr\},
\label{ham_coupl}
\ea
where we introduced two gap functions corresponding to two different regions of interactions $V_1$ and $V_2$
\beq
\Delta_j({\bf x})=V_j\langle \psi_{\downarrow}({\bf x})\psi_{\uparrow}({\bf x})\rangle.
\label{delta_x}
\eq

We proceed in the standard way \cite{abrikosov1975methods,fetter1971quantum}  in deriving the gap equation (see Supplemental Material), in some ways  similar to the case of a two-band superconductor \cite{Tilley64}. The final equations for $\Delta_1$ and $\Delta_2$ can be written in the following matrix form
\begin{equation}
  \begin{split}
      &\begin{bmatrix}
      \Delta_1 \\
      \Delta_2 
      \end{bmatrix}=
         \begin{bmatrix}
      (\lambda_1+\lambda_2)I_1 & \lambda_2I_2 \\
      \lambda_2 I_1 & \lambda_2I_2 \\
      \end{bmatrix}
      \begin{bmatrix}
      \Delta_1 \\
      \Delta_2
      \end{bmatrix},\\
      \end{split}
  \label{gap_equation}
\end{equation}
where 
\bea
I_1=\int_0^{\hbar\omega_B}  \frac{d\xi }{(\xi^2+\Delta_1^2)^{\frac{1}{2}}}\tanh\biggl[\frac{(\xi^2+\Delta_1^2)^{\frac{1}{2}}}{2k_BT}\biggr], \nn \\
I_2=\int_{\hbar \omega_B}^{\hbar\omega_D}  \frac{d\xi}{(\xi^2+\Delta_2^2)^{\frac{1}{2}}}\tanh\biggl[\frac{(\xi^2+\Delta_2^2)^{\frac{1}{2}}}{2k_BT}\biggr],
\ea
$\lambda_1=N(0)\gamma_1^2$ is the effective coupling constant due to bogolons, $\lambda_2=N(0)\gamma_2^2$ is the effective coupling constant due to phonons, $N(0)=mp_F/(2\pi^2)$ being the density of states on the Fermi surface in 2D. One should note that $\Delta_1$ and $\Delta_2$ are not two separate gaps, but just two constants which define one frequency dependent gap-function. 

{\it Critical temperature} The  critical temperature $T_c$ is obtained by linearizing the gap equation \eqref{gap_equation} by requiring  $T\rightarrow T_c$, $\Delta\rightarrow 0$  
\beq
\begin{vmatrix}  1-(\lambda_1+\lambda_2)I_1 & -\lambda_2 I_2 \\ -\lambda_2 I_1 & 1-\lambda_2I_2 \end{vmatrix}=0.
\label{det}
\eq
This is the determinant of the matrix in Eqs. \eqref{gap_equation} for the eigenvalue equal to unity. In the limit $T\rightarrow T_c$, $\Delta\rightarrow 0$  the integrals $I_1$ and $I_2$ can be expressed in terms of the digamma functions $\Psi$
\bea
I_1&=&\Psi\left(\frac{1}{2}+\frac{\omega_B}{2\pi T_c}\right)-\Psi\left(\frac{1}{2}\right)\approx \ln\left(\frac{2e^{\gamma}\omega_B}{\pi k_BT_c}\right),  \\
I_2&=&\Psi\left(\frac{1}{2}+\frac{\omega_D}{2\pi T_c}\right)-\Psi\left(\frac{1}{2}+\frac{\omega_B}{2\pi T_c}\right)\approx \ln\left(\frac{\omega_D}{\omega_B} \right), \nn
\label{lin_integrals}
\ea
here the approximate values are valid for $k_BT_c<<\hbar \omega_B<\hbar \omega_D$, and $\gamma$ is Euler's constant. The critical temperature is then 
\begin{equation}
k_BT_c\approx1.13{\hbar \omega_B} \exp\left({-\frac{1}{\lambda_1+\lambda^*_2}}\right),
\label{crit_temp}
\end{equation}
where $\lambda_2^*$ is the logarithmically renormalized (enhanced) interaction constant due to phonons:
\beq
\lambda^*_2=\frac{\lambda_2}{1-\lambda_2\ln\frac{\omega_D}{\omega_B}}.
\eq
One should note that in the limit $\hbar \omega_D>\hbar \omega_B>k_BT_c$, we would obtain similar to \eqref{crit_temp} equation but with $\omega_B\rightleftarrows \omega_D$, and $\lambda_1\rightleftarrows \lambda_2$.

\begin{figure}
\begin{center}
\includegraphics[width=0.5\textwidth]{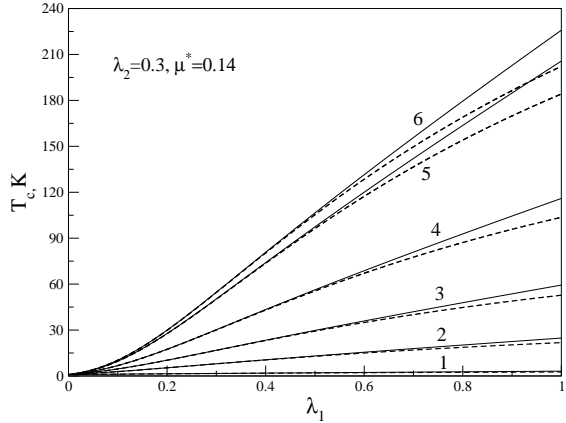}\vspace*{-0.8em} 
\end{center}
\caption{Critical temperature $T_c$ of the two-layered structure of aluminium superconductor ($\hbar\omega_D=428K, T_{c0}\approx 1K$) superimposed with pBEC calculated from the numerical solution of Eq. \eqref{tc_3} (solid lines). The critical temperature is shown versus the coupling constant arising due to the bogolon-induced attraction between electrons $\lambda_1$. $T_c$ is plotted for different $\omega_B$-s: $\hbar \omega_B=4.28 K$ (1),  $\hbar \omega_B=42.8K$ (2), $\hbar \omega_B=107K$ (3), $\hbar\omega_B=214K$ (4), $\hbar\omega_B=385.2 K$ (5), $\hbar\omega_B=423.72K$ (6). The dashed curved are obtained from Eq. \eqref{tc_3} with the integrals replaced by their approximate values \eqref{lin_integrals} for the corresponding $\omega_B$-s.  }
\label{numeric_tc}
\end{figure} 

We now estimate the effect of Coulomb interaction along the lines of  \cite{Bogoliubov58} for weak coupling (one could also do it as in \cite{Mcmillan68} for strong coupling theory). We extend the Gorkov equations for the case of three interacting constants (third constant $M$ being effective Coulomb interaction with the cut-off frequency $\omega_C=E_F/\hbar$). Introducing $\mu=N(0)M$ we get a similar to \eqref{det} equation (see also Supplemental Material, Section III)
\bea
1-(\lambda_1+\lambda_2)I_1-\lambda_2 I_2+\lambda_1\lambda_2 I_1I_2 \nn \\
+\mu^*(I_1+I_2)-\lambda_1\mu^* I_1 I_2=0,
\label{tc_3}
\ea
where $\mu^*=\mu/[1+\mu \ln(\omega_C/\omega_D)]$ is the logarithmically suppressed Coulomb interaction  as usual. We solve Eq. \eqref{tc_3} numerically as well as analytically, using the approximate expressions for the integrals \eqref{lin_integrals}. The analytical expression is  $k_BT_c\approx 1.13 \hbar\omega_B\exp\left(-1/(\lambda_1+\lambda_2-\mu^*) \right)$ and is valid in the limit $E_F>>\hbar\omega_D>>\hbar\omega_B>>k_BT_c$. In Fig. \ref{numeric_tc} we present the results for $T_c$ for fixed $\lambda_2, \mu^*$ and different $\omega_B$-s ($1\%, 10\%, 25\%, 50\%, 90\%$ and $ 99\%$ of $\hbar \omega_D$).  We take $\lambda_2\approx 0.3$ and $\mu^*\approx 0.14$ which approximately correspond to Al. 
 One should keep in mind that the validity range of the results is determined by the smallest frequency, in this case $\omega_B$. 

We see from Fig. \ref{numeric_tc} that even small $\lambda_1$ can lead to a substantial increase of the critical temperature $T_c$ in comparison to the bare one $T_{c0}$ provided the cut-off frequency for bogolons $\omega_B$ is of sizeable effect compared to $\omega_D$. 
Since both parameters ($\lambda_1$ and $\omega_B$) are tuneable, the set-up we suggest is very promising for obtaining superconductors with strongly enhanced critical temperature. We note that our simple model should be derivable from the strong coupling limit by approximating $\alpha^2 F(\omega) $ by two $\delta$-functions at $\omega_B$ and $\omega_D$. The procedure should result in further suppression of $T_c$ at large $\lambda$-s, but would however, not change our main conclusions. 

\begin{figure}
\begin{center}
\includegraphics[width=0.5\textwidth]{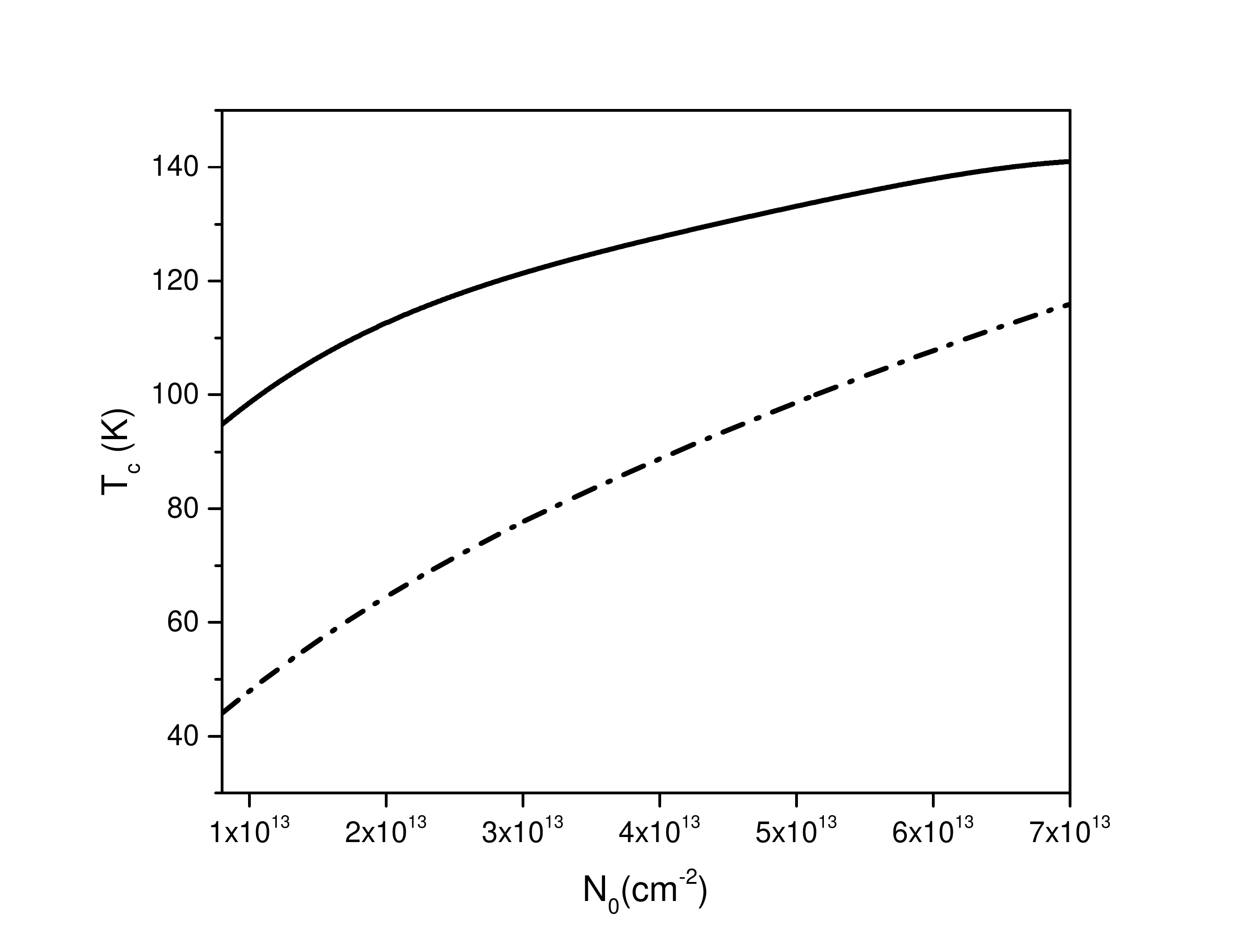}\vspace*{-0.8em} 
\end{center}
\caption{Critical temperature $T_c$ of the two-layered structure of aluminium superconductor ($\omega_D=429K$) superimposed with pBEC from the GaN layer. The critical temperature is shown versus polariton density $N_0$. Dashed curve represents the result derived by formula \ref{crit_temp}. Solid curve shows the result derived from numerical simulation of gap equation with two types of interaction potentials taken into account: electron-phonon, electron-polariton and Coulomb interaction in line with \cite{Cherotchenko2016,laussy2012polariton}. }
\label{crit_temp_Al}
\end{figure} 

In Fig. \ref{crit_temp_Al} we present the dependence of the critical temperature $T_c$ calculated by direct numerical solution of the gap equation exciton-polariton interaction potential taken from Ref.\cite{laussy2012polariton} with the additional phononic coupling (black curve), and by formula \eqref{crit_temp} (red curve)in the limit $\omega_B>\omega_D$ for a specific two-layered heterostructure, where superconducting layer is Aluminium sheet, while pBEC is induced in the GaN layer. Parameters used for calculation are taken from Ref.\cite{laussy2012polariton}. The difference between these results may come from not proper derivation of the exciton-polariton cut-off frequency $\omega_B$ in both cases. 

{\it{Conclusion}} We studied the superconducting critical temperature of a hybrid system where Cooper pairing is mediated by coupling to two types of excitations: the Bogoliubov excitations of the condensate and the  phononic excitations of the metal plate. We show that the additional coupling leads to a considerable enhancement of the critical temperature. 
We propose a concrete experimental  setup in which  superconductivity with the two couplings can be realized and estimate the critical temperature for specific realization of a Al superconductor coupled to a pBEC from the GaN layer.
Note, that the chosen GaN/AlGaN heterostructure allows for condensate stability up to the room temperatures due to the high exciton binding energy specific of the structure.  
Our model can be straightforwardly generalized to the case of multiband superconductors, such as pnictides, for instance, with our main conclusion about the dramatic increase of $T_c$ remaining qualitatively the same. For any specific structure a detailed calculation accounting for all terms in
the Hamiltonian (9) would be needed, however, the interplay between phonon and exciton superconductivity will remain important and will still result in the enhancement of $T_c$.

\end{document}